\documentclass[10pt,doublecolumn,twosides,]{IEEEtran}
\usepackage{ntheorem}
\usepackage{amsmath}
\usepackage{amsfonts}
\newtheorem{theorem}{Theorem}
\newtheorem{corollary}{Corollary}
\newtheorem{lemma}{Lemma}
\newtheorem{definition}{Definition}
\newtheorem{assumption}{Assumption}
\newtheorem{remark}{Remark}
\usepackage{ntheorem}
\usepackage{amsmath}
\usepackage{amsfonts}
\usepackage{listings}
\usepackage{color}
\usepackage{array}
\usepackage{tabu}
\usepackage{mathrsfs,amssymb}
\usepackage{graphicx}
\graphicspath{ {figure/} }
\usepackage{float}
\usepackage{algorithm}
\usepackage{booktabs}
{}

\usepackage{algpseudocode}
\algdef{SE}[DOWHILE]{Do}{doWhile}{\algorithmicdo}[1]{\algorithmicwhile\ #1}%
\DeclareMathOperator*{\argmin}{arg\,min}

\DeclareMathOperator*{\card}{card}
\DeclareMathOperator*{\supp}{supp}

\begin{document}
	
	\title{A Robust CACC Scheme Against Cyberattacks Via Multiple Vehicle-to-Vehicle Networks }
	
	%
	%
	%
	
\author{Tianci Yang \textsuperscript{1}, Carlos Murguia \textsuperscript{2}, Dragan Ne\v{s}i\'{c}\textsuperscript{3}, and Chen Lv\textsuperscript{1},
	\thanks{This work was supported by the SUG-NAP Grant (No. M4082268.050) of Nanyang Technological University, Singapore}
	\thanks{\textsuperscript{1} The authors are with the School of Mechanical and Aerospace Engineering, Nanyang Technological University, Singapore. Emails:
		\{tianci.yang, lyuchen\}@ntu.edu.sg}%
	\thanks{\textsuperscript{2} The author is with the Department of Mechanical Engineering, Eindhoven University of Technology, The Netherlands. Email:
		c.g.murguia@tue.nl}
	\thanks{\textsuperscript{3} The author is with the Department of Electrical and Electronic Engineering, University of Melbourne, Australia. Email:
		dnesic@unimelb.edu.au}%
}
	%
	%

	\markboth{Journal of \LaTeX\ Class Files,~Vol.~14, No.~8, April~2020}%
	{Shell \MakeLowercase{\textit{et al.}}: Bare Demo of IEEEtran.cls for IEEE Journals}
	%



	\maketitle
	\begin{abstract}
		Cooperative Adaptive Cruise Control (CACC) is a vehicular technology that allows groups of vehicles on the highway to form in closely-coupled automated platoons to increase highway capacity and safety, and decrease fuel consumption and CO2 emissions. The underlying mechanism behind CACC is the use of Vehicle-to-Vehicle (V2V) wireless communication networks to transmit acceleration commands to adjacent vehicles in the platoon. However, the use of V2V networks leads to increased vulnerabilities against faults and cyberattacks at the communication channels. Communication networks serve as new access points for malicious agents trying to deteriorate the platooning performance or even cause crashes. Here, we address the problem of increasing robustness of CACC schemes against cyberattacks by the use of multiple V2V networks and a data fusion algorithm. The idea is to transmit acceleration commands multiple times through different communication networks (channels) to create redundancy at the receiver side. We exploit this redundancy to obtain attack-free estimates of acceleration commands. To accomplish this, we propose a data-fusion algorithm that takes data from all channels, returns an estimate of the true acceleration command, and isolates compromised channels. Note, however, that using estimated data for control introduces uncertainty into the loop and thus decreases performance. To minimize performance degradation, we propose a robust $H_{\infty}$ controller that reduces the joint effect of estimation errors and sensor/channel noise in the platooning performance (tracking performance and string stability). We present simulation results to illustrate the performance of our approach.
	\end{abstract}
	
	\section{Introduction}
	During the past few decades, our society has been continuously confronted with problems of heavy traffic congestion caused by limited highway capacity. An effective way to increase road capacity is to decrease the inter-vehicle distance. Since this might threaten traffic safety if vehicles are human-driven, vehicle automation is required to guarantee safety. Adaptive Cruise Control (ACC) is a technology that adapts the velocity of vehicles to enforce a desired safe distance to preceding vehicles for platooning. However, ACC induce large inter-vehicle distances \cite{vahidi2003research} and amplifies disturbances (caused, e.g., by sudden acceleration variations of the lead vehicle) in the upstream direction of the platoon, i.e., ACC is string unstable \cite{naus2010string}. Cooperative ACC (CACC) solves these challenges by using Vehicle-to-Vehicle (V2V) wireless communication networks to transmit acceleration commands to adjacent vehicles in the platoon. The usage of V2V wireless communications plays an essential role for CACC to guarantee string stability \cite{naus2010string, Paper1996, Ploeg2014}, and increase traffic throughput \cite{van2006impact}. However, with the surge of vehicle connectivity, new security challenges have emerged as wireless vehicular networks increasingly serve as new access points for adversaries trying to disrupt the vehicle dynamics, see, e.g., \cite{El-rewini2020}\nocite{lurendeau2006threatsa}\nocite{car}\nocite{greenberg2015hackers}\nocite{petit2014potential}-\cite{amoozadeh2015security}, and references therein. In the so-called Jeep Cherokee attack \cite{greenberg2015hackers}, researchers were able to remotely stop the engine of the vehicle while it was driving down a busy highway. This forced Chrysler to issue a recall for 1.4 million vehicles of 7 different models leading to an immense financial burden. It is essential to realize that cyberattacks to network-connected vehicles pose a real threat to human life -- one vehicle hack could lead to catastrophic loss of life of not only the driver and passengers, but also pedestrians and other drivers. It follows that strategic mechanisms to identify and deal with cyberattacks on connected vehicles are a pressing need in this hyper-connected world.
	
	Most of the current literature on security of autonomous/cooperative vehicles focuses on cryptography-based solutions, i.e., it provides algorithmic results to design communication protocols for device authentication and data transmission, see, e.g.,  \cite{Chen2019,contreras2017internet}. There are only a few results focusing on minimizing the potential performance degradation induced by cyberattacks on network-connected vehicles. In \cite{Liu2019,Wyk2019}, the authors exploit sensor redundancy and provide detection and isolation algorithms for a single vehicle under sensor attacks. The problem of achieving vehicle-consensus in the presence of replay attacks is solved in \cite{Elizabeth2012}. The authors in \cite{biron2017resilient} provide a robust control scheme for platooning under Denial-of-Service (DoS) attacks. They achieve this by modelling DoS attacks as stochastic communication delays in the network. In \cite{merco2018replay}, a variety of algorithms to detect replay attacks in connected vehicles equipped with CACC are provided. In \cite{mousavinejad2019distributed}, the authors provide an algorithm for detecting cyberattacks on connected vehicles using set-membership filtering techniques. The problem of attack detection and state estimation for connected vehicles under sensor attacks is addressed in \cite{ju2020deception} using Unbiased Finite Impulse Response (UFIR) filters. An attack-resilient sensor fusion algorithm is proposed in \cite{ivanov2016attack}. The authors use the concept of \textit{abstract} sensors to obtain multiple measurements of the same physical variable. Each abstract sensor provides a set with all possible values of the true state. Then, an attack-free fused measurement is obtained by checking the intersections of these sets. In \cite{amoozadeh2015security}, the authors suggest switching from CACC to ACC as a potential countermeasure if any abnormal behavior is detected. Here, we provide an alternative solution by creating (and then exploiting) communication channel \textit{redundancy} in connected vehicles.
	
	In this manuscript, we address the problem of increasing robustness of CACC schemes against cyberattacks by using multiple Vehicle-to-Vehicle (V2V) wireless communication networks and a data fusion algorithm. The idea is to transmit acceleration commands multiple times through different communication networks to create redundancy at the receiver side. However, due to network-induced imperfections (e.g., channel noise and data dropouts) the received data from different channels, will, in general, be different even in the attack-free case. This makes it challenging to distinguish between healthy data coming from adjacent vehicles and corrupted data induced by cyberattacks. To overcome this challenge, we propose a data-fusion algorithm that takes data from all channels, returns an estimate of the true acceleration command, and isolates compromised channels. We provide estimation performance guarantees in terms of the level of uncertainty induced by the communication channels. We prove that our algorithm is guaranteed to work (i.e., it provides an attack-free estimate of acceleration commands), if less than half of the communication channels are under attack. This is a realistic assumption as the adversary may only extract the cypher keys of some of the channels \cite{dofe2015strengthening,wang2016against}, or the attacker may have limited resources. Note that even if the adversary is able to compromise all the channels, she/he may not be able to attack them all at the same time as the power supply is usually limited \cite{grover2014jamming}. We assume that the set of attacked channels is unknown and potentially \textit{time-varying}. Note, however, that using estimated acceleration commands for control introduces uncertainty into the loop and thus decreases performance. To minimize this performance degradation, we propose an robust $H_{\infty}$ controller that reduces the joint effect of estimation errors and sensor/channel noise in the platooning performance (tracking performance and string stability). In the presence of cyberattacks, we show that a separation principle between data fusion and control holds and the vehicle string can be stabilized by closing the loop with the fusion algorithm and the robust controller. We use Input-to-State Stability (ISS) \cite{sontag2008input} of the closed-loop system with respect to estimation errors to conclude stability of the tracking dynamics.
	
	The core of our fusion scheme is inspired by the work in \cite{Chong2015}, where the problem of state estimation for general \emph{continuous-time linear time-invariant (LTI) systems} is addressed. The authors propose a multi-observer estimator, using a bank of Luenberger observers, that provides a robust estimate of the system state in spite of sensor attacks. The main idea behind their estimation scheme is to place extra sensors in systems to create \emph{redundancy}. Exploiting sensor/actuator redundancy is now a standard technique in the research area of secure estimation and control in cyber-physical systems, see, e.g., \cite{Fawzi2014,Chong2015,Shoukry2017,Kim2016a, Yang2018a, yang2020multi}. We remark, however, that sensor/actuator redundancy can not protect connected vehicles from cyberattacks at inter-vehicle communication networks. Modern vehicles equipped with C-V2X communication tools are capable to communicate with each other via 3G/4G/5G and DSRC communication networks \cite{abboud2016interworking}, \cite{ghafari2014vehicular}. In these modern network channels interference can be neglected as different frequency bands are allocated \cite{lu2014connected, ezhilarasan2017review}. The latter enables the possibility to create network redundancy as multiple copies of the same data can be transmitted via different communication channels. To the best of the authors' knowledge, none of the existing results have taken advantage of the network redundancy in connected vehicles to improve their resilience to cyberattacks.
	
	The paper is organized as follows. In Section \ref{pre}, some preliminary results needed for the subsequent sections are presented. In Section \ref{sysd}, the considered vehicle platoon system is described. In Section \ref{estimation}, we show that our fusion scheme provides robust estimates of the transmitted signals despite attacks on vehicular networks. Algorithms for detecting and isolating attacked communication channels are presented in Section \ref{det}. The proposed robust CACC scheme and stability analysis are given in Section \ref{control}. Finally, in Section \ref{conclusion}, concluding remarks are given.
\section{Preliminaries}\label{pre}
\subsection{Notation}
	
We denote the set of natural numbers by $\mathbb{N}$, the symbol $\mathbb{R}$ stands for the real numbers, $\mathbb{R}_{>0}$ ($\mathbb{R}_{\geq 0}$) denotes the set of positive (non-negative) real numbers, and $\mathbb{R}^{n\times m}$ is the set of all $n\times m$ real matrices, $m,n \in \mathbb{N}$. For any vector $v\in\mathbb{R}^{n}$,  we denote {$v^{J}$} the stacking of all $v_{i}$, $i\in J$, $J\subset \left\lbrace 1,\hdots,n\right\rbrace$, $|v|=\sqrt{v^{\top} v}$, $||v||_{\mathcal{L}_{p}}$ the signal $\mathcal{L}_p$-norm, and $\supp(v)=\left\lbrace i\in\left\lbrace 1,\hdots,n\right\rbrace |v_{i}\neq0\right\rbrace $. We denote the cardinality of a set $S$ as $\card(S)$. The binomial coefficient is denoted as $\binom{a}{b}$, where $a,b$ are non-negative integers. We denote a variable $m$ uniformly distributed in the interval $(z_{1},z_{2})$ as $m\sim\mathcal{U}(z_{1},z_{2})$ and normally distributed with mean $\mu$ and variance $\sigma^2$ as $m\sim \mathcal{N}(\mu,\sigma^2)$. The $n \times m$ matrices composed of only ones and only zeros are denoted by $\mathbf{1}_{n \times m}$ and $\mathbf{0}_{n \times m}$, respectively, or simply $\mathbf{1}$ and $\mathbf{0}$ when their dimensions are clear.
\begin{definition}[Vehicle String Stability]\emph{\cite{Ploeg2011}}
		Consider a string of $m\in \mathbb{N}$ interconnected vehicles. The string is said to be string stable
		if and only if
		\begin{equation}
			\begin{split}
				||z_{i}(t)||_{\mathcal{L}_{p}} \leq ||z_{i-1}(t)||_{\mathcal{L}_{p}},\hspace{1mm} \forall \hspace{2mm} t\geq 0, \hspace{1mm} 2 \leq i \leq m,	
			\end{split}
		\end{equation}
		where $z_{i}(t)$ can either be the inter-vehicle tracking error $e_{i}(t)$, the velocity
		$v_{i}(t)$, or the acceleration $a_{i}(t)$ of the $i$-th vehicle; $z_{i}(0) = 0$, $2 \leq i \leq m$; and $z_{1}(t)\in\mathcal{L}_{p}$ is an unknown input signal (e.g., acceleration commands of the lead vehicle in the string or disturbances). That is, a vehicle string is said to be string stable if inputs driving the lead vehicle (acceleration commands and/or disturbances) do not amplify throughout the string.
	\end{definition}

	\section{System Description}\label{sysd}
	\begin{figure}[t]\centering
		\includegraphics[width=0.5\textwidth]{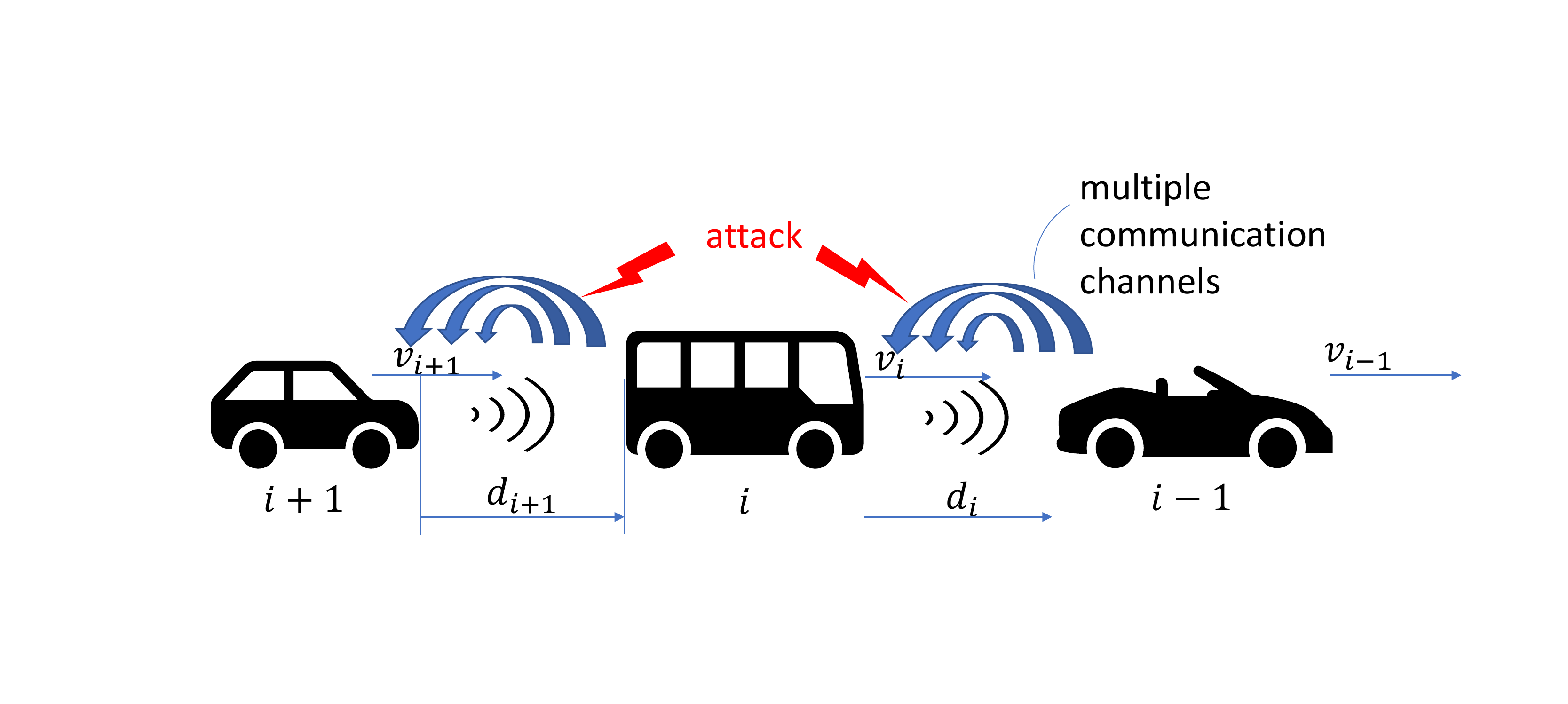}
		\caption{CACC-equipped vehicle platoon under network attacks: each vehicle is equipped with multiple communication channels to obtain information from its preceding vehicle. Note that some channels are under attack.}
		\centering
		\label{fig:1}
	\end{figure}
	Consider a platoon of $m$ vehicles as depicted in Figure \ref{fig:1}. Let $v_{i} \in \mathbb{R}$ denote the velocity of vehicle $i$, and $d_{i} \in \mathbb{R}$ be the distance between vehicle $i$ and its preceding vehicle $i-1$. The objective of each vehicle is to follow the preceding vehicle at the desired distance:
	\begin{equation}
		d_{r,i}(k) := r_{i}+h_{i}v_{i}(k), \hspace{1mm} i\in S_{m},
	\end{equation}
	where $h_{i} \in \mathbb{R}_{>0}$ denotes the constant head time spacing policy of the $i-$th vehicle, and $r_{i}$ is the standstill distance. Define $S_{m} := \left\lbrace i\in\mathbb{N} | 1\leq i\leq m\right\rbrace $ as the set of all vehicles in a platoon of length $m \in \mathbb{N}$. Including the spacing policy term, $h_{i}v_{i}(k)$, in the desired inter-vehicle distance is known to improve string stability \cite{Paper1996, Ploeg2014}. The spacing error $e_{i} \in \mathbb{R}$ is then defined as
	\begin{equation}\label{1}
		\begin{split}
			e_{i}(k) := &d_{i}(k)-d_{r,i}(k),\\
			=&(q_{i-1}(k)-q_{i}(k)-L_{i})-(r_{i}+h_{i}v_{i}(k)),
		\end{split}
	\end{equation}
	with $q_{i} \in \mathbb{R}$ being the rear-bumper position of vehicle $i$ and $L_{i}$ its length. It has been proved that the dynamic controller introduced in \cite{Ploeg2011} accomplishes the vehicle-following objective, $\lim_{k\to\infty}e_{i}(k)=0, \forall i\in S_{m}$, and guarantees string stability. Here, we propose a switching controller, inspired by the one in \cite{Ploeg2011}, that robustifies the platoon against cyberattacks.

As a basis for the controller design, the following vehicle model is adopted \cite{Ploeg2011}:
	\begin{equation}\label{2}
		\begin{split}
			\begin{bmatrix}
				\dot{d}_{i}\\
				\dot{v}_{i}\\
				\dot{a}_{i}
			\end{bmatrix}=\begin{bmatrix}
			v_{i-1}-v_{i}\\a_{i}\\\frac{1}{\tau_{i}}a_{i}+\frac{1}{\tau_{i}}u_{i}		\end{bmatrix}, \hspace{1mm} i\in S_{m},
		\end{split}
	\end{equation}
	where $a_{i}$ is the acceleration of vehicle $i$, $u_{i}$ is the vehicle controller (the desired acceleration), and $\tau_{i}$ is a time constant representing the driveline dynamics. Consider the following dynamic controller:
	\begin{equation} \label{3a}
		h\dot{u}_{i}=-u_{i}+\xi_{i},
	\end{equation}
with 
\begin{equation}\label{3}
\begin{split}
	\xi_{i}: = K_{i}\begin{bmatrix}
		e_{i}\\
		\dot{e}_{i}\\
		\ddot{e}_{i}
	\end{bmatrix}+ \hat{u}_{i-1},\hspace{3mm}i\in S_{m},
\end{split}
\end{equation}
$K_{i} := \begin{bmatrix}
	k_{pi}&k_{di}&k_{ddi}
\end{bmatrix}$, and $k_{ddi}=0$ (to avoid feedback of jerk). The term $\hat{u}_{i-1}$ denotes an estimate of the control input of the preceding vehicle, $u_{i-1}$. We compute this estimate in real-time using multiple distorted copies of $u_{i-1}$ (transmitted wirelessly by vehicle $i-1$ via multiple communication networks). In \cite{Ploeg2011}, the authors assume that $u_{i-1}$ can be transmitted directly to vehicle $i$ free of noise and network effects. However, even in that idealized scenario, if $u_{i-1}$ is tampered with by cyberattacks, safety cannot be guaranteed. Here, we robustify the controller by transmitting multiple distorted copies of $u_{i-1}$ (to create redundancy and thus not depending on a single, possibly corrupted, signal) and using this redundant information to construct an estimate $\hat{u}_{i-1}$ of $u_{i-1}$. We use $\hat{u}_{i-1}$ as a feedforward term to drive the controller in \eqref{3}. By doing so, we do not rely on a single signal but we pay the price of working with estimates (thus introducing uncertainty) and having to devise a fusing algorithm that produces accurate approximations of $u_{i-1}$ and ensure stability of the platoon in closed-loop with controller \eqref{3}.

Using \eqref{1}-\eqref{3}, the following platoon model is obtained:
	\begin{equation}\label{s1}
		\begin{split}
			\begin{bmatrix}
				\dot{e}_{i}\\
				\dot{v}_{i}\\
				\dot{a}_{i}\\
				\dot{u}_{i}
			\end{bmatrix}=&\begin{bmatrix}
				0&-1&-h_{i}&0\\
				0&0&1&0\\
				0&0&\frac{-1}{\tau_{i}}&\frac{1}{\tau_{i}}\\
				\frac{k_{pi}}{h_{i}}&\frac{-k_{di}}{h_{i}}&-k_{di}&\frac{-1}{h_{i}}
			\end{bmatrix}\begin{bmatrix}
				{e}_{i}\\
				{v}_{i}\\
				{a}_{i}\\
				{u}_{i}
			\end{bmatrix}\\
			&+\begin{bmatrix}
				1&0\\
				0&0\\
				0&0\\
				\frac{k_{di}}{h_{i}}&\frac{1}{h_{i}}
			\end{bmatrix}\begin{bmatrix}
				v_{i-1}\\
				\hat{u}_{i-1}
			\end{bmatrix}.
		\end{split}
	\end{equation}
	 Without loss of generality, we assume $r_{i}$ is equal to zero. Here, we assume $v_{i-1}$ is obtained from on-board sensors of vehicle $i$ that measure the velocity of vehicle $i$ and the relative velocity between vehicles $i$ and $i-1$. We rewrite system \eqref{s1} compactly in matrix form as follows:
	\begin{equation}
		\begin{split}
			\dot{x}_{i}=A_{ci}x_{i}+B_{ci}\epsilon_{i},
		\end{split}
	\end{equation}
	with $x_{i} := [
		e_{i} \hspace{1mm} v_{i} \hspace{1mm} a_{i} \hspace{1mm} u_{i}
	]^{\top}$, $\epsilon_{i} := [
		v_{i-1} \hspace{1mm} \hat{u}_{i-1}]^{\top}$, and matrices $A_{ci}$ and $B_{ci}$ defined according to \eqref{s1}.
	
	We assume the first vehicle in the platoon (the leader) does not have a preceding vehicle. Instead,
	it follows a so-called \emph{virtual reference vehicle} ($i = 0$), allowing the lead vehicle to employ the same controller as the other vehicles in the platoon. We rewrite the virtual vehicle dynamics in matrix form:
	\begin{equation}\label{s2}
		\begin{split}
			\begin{bmatrix}
				\dot{e}_{0}\\
				\dot{v}_{0}\\
				\dot{a}_{0}\\
				\dot{u}_{0}
			\end{bmatrix}=\begin{bmatrix}
				0&0&0&0\\
				0&0&1&0\\
				0&0&-\frac{1}{\tau_{0}}&\frac{1}{\tau_{0}}\\
				0&0&0&-\frac{1}{h_{0}}
			\end{bmatrix}	\begin{bmatrix}
				{e}_{0}\\
				{v}_{0}\\
				{a}_{0}\\
				{u}_{0}
			\end{bmatrix}+\begin{bmatrix}
				0\\0\\0\\\frac{1}{h_{0}}
			\end{bmatrix}\epsilon_{0},
		\end{split}
	\end{equation}
	where $\epsilon_{0}$ is the desired acceleration of the leader induced by the human driver. We can write the virtual vehicle dynamics in matrix form as follows:
	\begin{equation}
		\begin{split}
			\dot{x}_{0}=A_{c0}x_{0}+B_{c0}\epsilon_{0},
		\end{split}
	\end{equation}
	with $x_{0} := [
		e_{0} \hspace{1mm} v_{0} \hspace{1mm} a_{0} \hspace{1mm} u_{0} ]^{\top}$, and matrices $A_{c0}$ and $B_{c0}$ defined from \eqref{s2}.
	
We exactly discretize \eqref{s1} and \eqref{s2} at the sampling time instants, $t = T_{s}k$, $k \in \mathbb{N}$, and assume a zero-order hold to implement control actions (see \cite{Astrom} for details) to obtain the following equivalent discrete-time systems:
	\begin{equation}\label{d1}
		x_{i}(k+1)=A_{i}^dx_{i}(k)+B_{i}^d\epsilon_{i}(k), \hspace{1mm} i\in S_{m},
	\end{equation}
	and
	\begin{equation}\label{d2}
		x_{0}(k+1)=A_{0}^dx_{0}(k)+B_{0}^d\epsilon_{0}(k),
	\end{equation}
	with $x_{i}(k) := x_{i}(T_{s}k)$, $\epsilon_{i}(k) := \epsilon_{i}(T_{s}k)$, $k \in \mathbb{N}$, $A_{i}^d = e^{A_{i}^cT_{s}}$, and $B_{i}^d=\left( \int_{0}^{T_{s}}e^{A_{i}^c(T_{s}-s)}ds\right) B_{i}^c$, for $i\in S_{m}\cup \left\lbrace 0\right\rbrace$.

	\section{Secure data fusion for connected vehicles equipped with multiple networks}\label{estimation}
In this section, we propose a fusion algorithm that exploits communication channel redundancy in connected vehicles to construct a robust estimate $\hat{u}_{i-1}$ of the preceding vehicle's acceleration commands. We assume the desired acceleration of vehicle $(i-1)$, $u_{i-1}$, for $i\in S_{m}\setminus \left\lbrace 1\right\rbrace $, is transmitted from vehicle $i-1$ to vehicle $i$ through $N_{i} \geq 3$ different communication channels, i.e., at each time $k\geq 0$, vehicle $i$ receives $N_{i}$ distorted copies of $u_{i-1}(k)$:
	\begin{equation}\label{sss}
		\begin{split}
			U_{i}(k) &= \left[ \begin{matrix}
				U_{i1}(k)\\
				U_{i2}(k)\\
				\vdots\\
				U_{iN_{i}}(k)
			\end{matrix}\right] := \left[ \begin{matrix}
				u_{i-1}(k)+\nu_{i1}(k) + \eta_{i1}(k)\\
				u_{i-1}(k)+\nu_{i2}(k) + \eta_{i2}(k) \\
				\vdots\\
				u_{i-1}(k)+\nu_{iN_{i}}(k) + \eta_{iN_i}(k)
			\end{matrix}\right] \\
		&=: \mathbf{1}u_{i-1}(k) + \nu_i(k) + \eta_i(k),
		\end{split}
	\end{equation}
	where $\nu_{ij}\in\mathbb{R}$, $\left\lbrace \nu_{ij}(k)\right\rbrace \in l_{\infty}$, denotes \emph{unknown} additive perturbations in the $j$-th channel, and $\eta_{ij}\in\mathbb{R}^{N}$ is the potentially unbounded attack signal, $j\in\left\lbrace 1,\ldots,N_{i}\right\rbrace $. That is, if the $j$-th channel is compromised, then $\eta_{ij}(k)\neq 0$ for some $k\geq 0$; otherwise, $\eta_{ij}(k)=0$ for all $k\geq 0$. We let $W_{i}(k)\subset\left\lbrace 1,\ldots,N_{i}\right\rbrace $ be the set of attacked communication channels between vehicles $i$ and $i-1$ at time $k$, i.e.,
	\begin{equation}
		\supp(\eta_{i}(k))\subseteq W_{i}(k).
	\end{equation}
The vector of perturbations, $\nu_{i} := (\nu_{i1},\ldots,\nu_{iN}) \in \mathbb{R}^{N_i}$, encompass all network induced imperfections in the channels (e.g., noise, delays, packet dropouts, and quantization). In \eqref{sss}, we are implicitly assuming that the adversary can attack communication channels by additively injecting an arbitrary vector $\eta_{i} := (\eta_{i1},\ldots,\eta_{iN}) \in \mathbb{R}^{N_i}$.

\begin{remark}
Different from existing results in the literature, e.g., \emph{\cite{Chong2015,Fawzi2014a,Showkatbakhsh2017,Yang2018a,Shoukry2017,Kim2016a,yang2020multi}}, which assume the set of attacked nodes is time-invariant, here we slightly relax the assumption by allowing the set of attacked communication channels to be time-varying, i.e., the attacker can choose a different set of communication channels to compromise at each time $k\geq 0$, \emph{\cite{grover2014jamming,mpitziopoulos2007defending,muraleedharan2006jamming}}.
\end{remark}
To implement the CACC controller \eqref{3a}-\eqref{3}, we need the estimate $\hat{u}_{i-1}$ of the acceleration $u_{i-1}$ of the preceding vehicle. What vehicle $i$ receives is the vector $U_{i}(k)$ containing $N_{i}$ corrupted copies of $u_{i-1}(k)$ (corrupted by network-induced perturbations and cyberattacks). Here, we seek to reconstruct $u_{i-1}(k)$ from $U_{i}(k)$ and implement the controller \eqref{3a}-\eqref{3} that uses the reconstructed signal, $\hat{u}_{i-1}(k)$. Then, the first question to raise is whether $u_{i-1}(k)$ is actually \emph{reconstructible} from $U_{i}(k)$ (in some appropriate sense).
\begin{definition}
Consider input $u_{i-1}(k)$ and the vector of distorted inputs $U_i(k)$ in \eqref{sss} under $q_i$ attacks $\eta_i(k)$, $\supp(\eta_{i}(k))\subseteq W_{i}(k)$. Input $u_{i-1}(k)$ is reconstructible from $U_i(k)$, if for every other input $\bar{u}_{i-1}(k)$ with corresponding $\bar{U}_i(k)$ under $q_i$ attacks and attack vector $\bar{\eta}_i(k)$, $\supp(\bar{\eta}_{i}(k)) \subseteq \bar{W}_{i}(k)$, the following is satisfied: 
\begin{equation}
U_i(k)=\bar{U}_i(k)\Longrightarrow u_{i-1}(k)=\bar{u}_{i-1}(k).
\end{equation}
\end{definition}
Definition 1 implies that for $u_{i-1}(k)$ and corresponding $U_i(k)$, there is no other $\bar{u}_{i-1}(k) \neq u_{i-1}(k)$ consistent with the received $U_i(k)$. That is, $u_{i-1}(k)$ is reconstructible if it can be uniquely determined from $U_i(k)$.

\begin{theorem}\label{b}For every integer $q_{i}\geq 0$, the following statements are equivalent:
	
		(i) $u_{i-1}(k)$ is reconstructible from $U_{i}(k)$ under $q_{i}$ attacks for $k\geq 0$.
	
	(ii) $q_{i}<\frac{N_{i}}{2}$.
\end{theorem}
\textit{Proof:} See Appendix \ref{the1}. \hfill$\blacksquare$
\vspace{3mm}

From Theorem 1, it follows that the following assumption is needed for any reconstruction scheme to work.

	\begin{assumption}\label{a1}
		The number of attacked channels between vehicle $i$ and $(i-1)$ does not exceed $\dfrac{N_{i}}{2}$, i.e.,
		\begin{equation}
			\card(W_{i}(k))\leq q_{i}<\frac{N_{i}}{2}.
		\end{equation}
	\end{assumption}
	\begin{remark}
	Note that Assumption \ref{a1} allows for attacks on all the channels as long as, at every time-step, the number of injected attack signals are strictly less than $\dfrac{N_{i}}{2}$. That is, the set of compromised channels can be time-varying as long as a maximum of $q_{i} < \dfrac{N_{i}}{2}$ channels are attacked at every $k \in \mathbb{N}$. This limitation might result from hardware or energy constraints from the adversary point of view. For instance, a jamming attack may randomly select one channel to jam for a short period of time \emph{\cite{mpitziopoulos2007defending}}; and channel-hopping and pulsed-noise jammers might jam different sets of channels at different time instants \emph{\cite{grover2014jamming,muraleedharan2006jamming}}.
	\end{remark}
	\begin{corollary}\label{c2}
		Under Assumption \ref{a1}, among all $N_{i}$ communication channels, at least $N_{i}-q_{i}$ of them are attack-free; and among every set of $N_{i}-q_{i}$ communication channels, at least $N_{i}-2q_{i}$ channels are attack-free.\\
\noindent \emph{Corollary \ref{c2} follows trivially from Assumption 1.}
\end{corollary}

\subsection{Reconstruction Strategy}

Before we present our reconstruction algorithm, we need to introduce some notation and mathematical machinery. For every subset $J\subset\left\lbrace 1, \ldots, N_{i}\right\rbrace $ of communication channels and $k \geq 0$, define $\hat{u}_{iJ}(k)$ as the average value of all the data transmitted via $J$ channels at time $k$:
	\begin{equation}
		\hat{u}_{iJ}(k) :=  \frac{\sum_{j\in J}U_{ij}(k)}{\card(J)};
	\end{equation}
and an (unknown) upper bound on all channel disturbances between vehicles $i$ and $(i-1)$:
	\begin{equation}
		||\nu_{i}||_{\infty} := \max_{j\in\left\lbrace 1,\cdots,N_i\right\rbrace }\left\lbrace ||\nu_{ij}||_{\infty}\right\rbrace.
	\end{equation}
	\begin{lemma}
		If $\eta_{i}^{J}(k)= \mathbf{0}$, then
		\begin{equation}
			|\hat{u}_{iJ}(k)-u_{i-1}(k)|\leq ||\nu_{i}||_{\infty},
		\end{equation}
		for all $k\geq 0$.
	\end{lemma}
	\textit{Proof:}
See Appendix \ref{l1}.
	\hfill$\blacksquare$
\vspace{3mm}	

Under Assumption \ref{a1}, there exists at least one subset $\bar{I}(k)\subset \left\lbrace 1, \ldots, N_{i}\right\rbrace $ with $\card(\bar{I}(k))=N_{i}-q_{i}$ such that $\eta_{i}^{\bar{I}(k)}(k)= \mathbf{0}$ for $k \geq 0$. Then, in general, the difference between $\hat{u}_{\bar{I}(k)}(k)$ and any $U_{ij}(k)$ (see \eqref{sss}), $i\in\bar{I}(k)$, will be less than the other subsets $J\subset\left\lbrace 1, \ldots, N_{i}\right\rbrace $ with $\card(J)=N_{i}-q_{i}$ and $\eta_{i}^{J}(k)\neq \mathbf{0}$. This motivates the following reconstruction algorithm.
	
\begin{figure}[t]\centering
 	\includegraphics[width=0.5\textwidth]{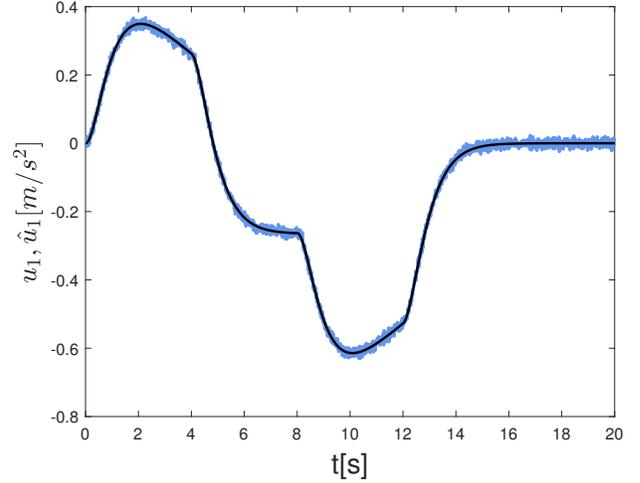}
 	\caption{$u_{i-1}$ (black) and $\hat{u}_{i-1}$ (blue).}
 	\centering
 	\label{fig:fs}
 \end{figure}

For every subset $J\subset\left\lbrace 1,\ldots,N_{i}\right\rbrace $ of channels with $\card(J)=N_{i}-q_{i}$, define $\pi_{iJ}(k)$ as the largest difference between $\hat{u}_{iJ}(k)$ and $U_{ij}(k)$ for all $j\in J$, i.e.,
	\begin{equation}\label{es1}
		\pi_{iJ}(k) :=\underset{j\in J}{\max}\left| \hat{u}_{iJ}(k)-U_{ij}(k)\right|,
	\end{equation}
	for all $k\geq 0$, and the sequence $\sigma_{i}(k)$ as
	\begin{equation}\label{es2}
		\begin{split}
			\sigma_{i}(k) := \underset{J\subset\left\lbrace 1,\ldots, N_{i}\right\rbrace: \card(J)=N_{i}-q_{i}}{\argmin}\pi_{iJ}(k).
		\end{split}
	\end{equation}
	Then, as proved below, the fused measurement indexed by $\sigma_{i}(k)$:
	\begin{equation}\label{es3}
		\hat{u}_{i-1}(k)=\hat{u}_{\sigma_{i}(k)}(k),
	\end{equation}
	is an attack-free measurement of $u_{i-1}(k)$. The following result uses the terminology presented above.
	\begin{theorem}\label{t2}
		Consider system \eqref{s1}, and the reconstruction algorithm \eqref{es1}-\eqref{es3}. Define the reconstruction error $e_{\sigma_{i}(k)}(k):=\hat{u}_{\sigma_{i}(k)}(k)-u_{i-1}(k)$, and let Assumption \ref{a1} be satisfied; then,
		\begin{equation}
			|e_{\sigma_{i}(k)}(k)|\leq 3||\nu_{i}||_{\infty}\label{sa},
		\end{equation}
		for all $k\geq 0$.
	\end{theorem}
	\textit{Proof:} See Appendix \ref{the2}.
 \hfill $\blacksquare$ \\[2mm]
	\textbf{Example 1.} We consider a platoon of $2$ vehicles. We let $h_{1}=h_{2}=0.5$, $\tau_{1}=\tau_{2}=0.1$. Exact discretization method is adopted with $T_{s}=0.01$ seconds. Let $N_{2}=3$, i.e., $3$ communication channels are used with corresponding channel noises $\nu_{21}\sim\mathcal{U}(-0.01,0.01)$, $\nu_{22}\sim\mathcal{U}(-0.02,0.02)$, and $\nu_{23}\sim\mathcal{U}(-0.03,0.03)$. We use sensor noise $\omega_{d},\omega_{v}\sim\mathcal{U}(-0.1,0.1)$ and let $\epsilon_{0}$ be given as in Table 1.
\begin{table}[h!]
	\centering
			\begin{tabular}{c|c}
				\hline
		$\epsilon_{0}(m/s^{2})$& Time(seconds)\\
		\hline
		-10&[0,5]\\
		0&(5,10]\\
		-10&(10,15]\\
		0&(15,20]\\
		\hline
		\end{tabular}
	\vspace{1mm}
		\label{tab:Table1}
		\caption{Desired acceleration of the leader.}
\end{table}
 At each time step $k$, let one of the $3$ communication channels between the two vehicles be randomly selected to be attacked, i.e., $W_{2}(k)=\left\lbrace 1\right\rbrace$, $\left\lbrace 2\right\rbrace $ or $\left\lbrace 3\right\rbrace $ and let $\eta_{W_{2}(k)}\sim\mathcal{N}(0,5^{2})$. For all $k\geq 0$, the second vehicle uses \eqref{es1}-\eqref{es3} to construct an estimate $\hat{u}_{1}(k)$ of $u_{1}(k)$. The performance of the fusion algorithm is shown in Figure \ref{fig:fs}.
	\begin{section}{Detection and isolation}\label{det}
		In this section, we now assume bounds on the perturbations in all channels are known, i.e., $||\nu_{ij}||_{\infty}$ for $j\in\left\lbrace 1, \ldots, N_{i}\right\rbrace $, and $i\in S_{m} \setminus \left\lbrace 1\right\rbrace $ is known. We first provide a simple technique for detecting attacks on the communication channels. Then, we use fusion algorithm presented in Section \ref{estimation} to select the attack-free channels and isolate the ones that are compromised.

		\subsection{Detection strategy}
		If all the channels are attack-free, then the deviation between $\frac{\sum_{j=1}^{N_{i}}U_{ij}(k)}{N_{i}}$ and $U_{ij}(k)$, for all $j$, will be small and
		\begin{equation}\label{d11}
			\begin{split}
				&\left| \frac{\sum_{j=1}^{N_{i}}U_{ij}(k)}{N_{i}}-U_{ij}(k)\right|\\
				\leq&\left| \frac{\sum_{j=1}^{N_{i}}U_{ij}(k)}{N_{i}}-u_{i-1}(k)\right|+|\nu_{ij}(k)
				|\\ \leq&||\nu_{i}||_{\infty}+||\nu_{ij}||_{\infty}, \forall i\in\left\lbrace 1,\ldots,N_{i}\right\rbrace.
			\end{split}
		\end{equation}
		Define
		\begin{equation}\label{d12}
			\tau_{ij}^d := ||\nu_{i}||_{\infty}+||\nu_{ij}||_{\infty},
		\end{equation}
		for $j\in\left\lbrace 1,\ldots, N_{i}\right\rbrace $. Then, attacks are detected at time $k$ if there exists at least one channel $j\in\left\lbrace 1, \ldots, N_{i}\right\rbrace $ such that
		\begin{equation}\label{d13}
			\left| \frac{\sum_{j=1}^{N_{i}}U_{ij}(k)}{N_{i}}-U_{ij}(k)\right|>\tau_{ij}^d,
		\end{equation}
		for some $k\geq 0$.
		\subsection{Isolation strategy}
		From Section \ref{estimation}, we know that $\sigma_{i}(k)\subset \left\lbrace 1, \ldots, N_{i}\right\rbrace $ is a set of attack-free communication channels. For $t\geq 0$, we randomly select one channel $j^{*}(k)\in\sigma_{i}(k)$ and the corresponding $\eta_{ij^{*}(k)}(k)$ must satisfy $\eta_{ij^{*}(k)}(k)= 0$. For each $j\in\left\lbrace 1, \ldots, N_{i}\right\rbrace $ and $k \geq 0$, we compute the difference between $U_{ij}(k)$ and $U_{ij^{*}(k)}(k)$. Then, if the $j$-th channel is attack-free, i.e., $\eta_{ij}(k)=0$, we have
		\begin{equation}
			\begin{split}
				&|U_{ij^{*}(k)}(k)-U_{ij}(k)|\\=&|U_{ij^{*}(k)}(k)-u_{i-1}(k)+u_{i-1}(k)-U_{ij}(k)|\\
				\leq&|U_{ij^{*}(k)}(k)-u_{i-1}(k)|+|u_{i-1}(k)-U_{ij}(k)|\\
				\leq&||\nu_{ij^{*}(k)}||_{\infty}+||\nu_{ij}||_{\infty}.
			\end{split}
		\end{equation}
		For each $j\in\left\lbrace 1, \ldots, N_{i}\right\rbrace $, define
		\begin{equation}\label{i1}
			\begin{split}
				\tau_{ij}(k) := ||\nu_{ij^{*}(k)}||_{\infty}+||\nu_{ij}||_{\infty}.
			\end{split}
		\end{equation}
		Then, for $k\geq 0$, the $j$-th channel is isolated as an attacked one if
		\begin{equation}\label{i2}
			|U_{ij^{*}(k)}(k)-U_{ij}(k)|>\tau_{ij}(k).
		\end{equation}
		Then, the set of channels that is isolated as the attacked ones at time $k$, which we denote as $\hat{W}_{i}(k)$, is given as
		\begin{equation}\label{i3}
			\begin{split}
				\hat{W}_{i}(k) := \left\lbrace i\in\left\lbrace 1, \ldots, N_{i}\right\rbrace \left|  |U_{ij^{*}(k)}(k)-U_{ij}(k)|>\tau_{ij}(k)\right.\right\rbrace .
			\end{split}
		\end{equation}
	\vspace{3mm}\\
		\textbf{Example 2.} We consider a vehicle platoon of $2$ vehicles. We let $h_{1}=0.6$, $h_{2}=0.5$, $\tau_{1}=\tau_{2}=0.1$. Exact discretization method is adopted with $T_{s}=0.01$ seconds. Let $N_{2}=3$, i.e., $3$ communication channels are used with corresponding channel noises $\nu_{21}\sim\mathcal{U}(-0.1,0.1)$, $\nu_{22}\sim\mathcal{U}(-0.2,0.2)$, $\nu_{23}\sim\mathcal{U}(-0.3,0.3)$. Let $\epsilon_{0}$ be given as in Table 1. At each time step $k$, let one of the $3$ communication channels between the two vehicles be randomly selected to be attacked, i.e., $W_{2}(k)=\left\lbrace 1\right\rbrace$, $\left\lbrace 2\right\rbrace $ or $\left\lbrace 3\right\rbrace $ and let $\eta_{W_{2}(k)}\sim\mathcal{N}(0,5^{2})$. Then, $||\nu_{2}||_{\infty}=0.3$, and $\tau_{d21}=0.4$, $\tau_{d22}=0.5$, $\tau_{d23}=0.6$ accordingly. For $k\in[1,400]$, \eqref{d12}-\eqref{d13} are used for attack detection and it turns outs that $371$ out of $400$ time steps our detection algorithm successfully detects the attacks. For $k\in[1,20]$, \eqref{i1}-\eqref{i3} are used for isolating the attacked channels. The performance of our isolation algorithm is presented in Figure \ref{fig:ff}, where it is shown that our algorithm  successfully isolates the attacked channel $14$ out of $20$ time steps for the considered time-varying attacks.
		\vspace{3mm}
			\begin{remark}
			Note that \eqref{d13} and \eqref{i2} are only sufficient conditions for the detection and isolation algorithms to work respectively. This might lead the proposed algorithms conservative due to the perturbations in the communication channels. Example 2 illustrates that the algorithms still detect and isolate attacked channels relatively well under reasonable conditions.
		\end{remark}
		\begin{figure}[t]\centering
			\includegraphics[width=0.5\textwidth]{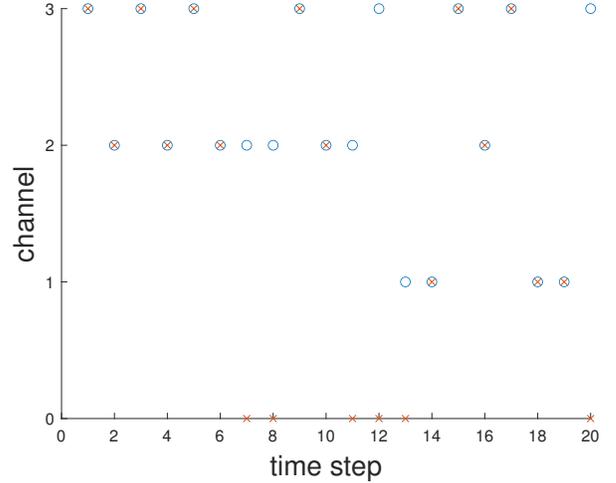}
			\caption{The actual attacked channel ('o') and the isolated channel ('x').}
			\centering
			\label{fig:ff}
		\end{figure}
	\end{section}
	\section{Robust controller and secure estimator}\label{control}
	\begin{figure}[t]\centering
		\includegraphics[width=0.5\textwidth]{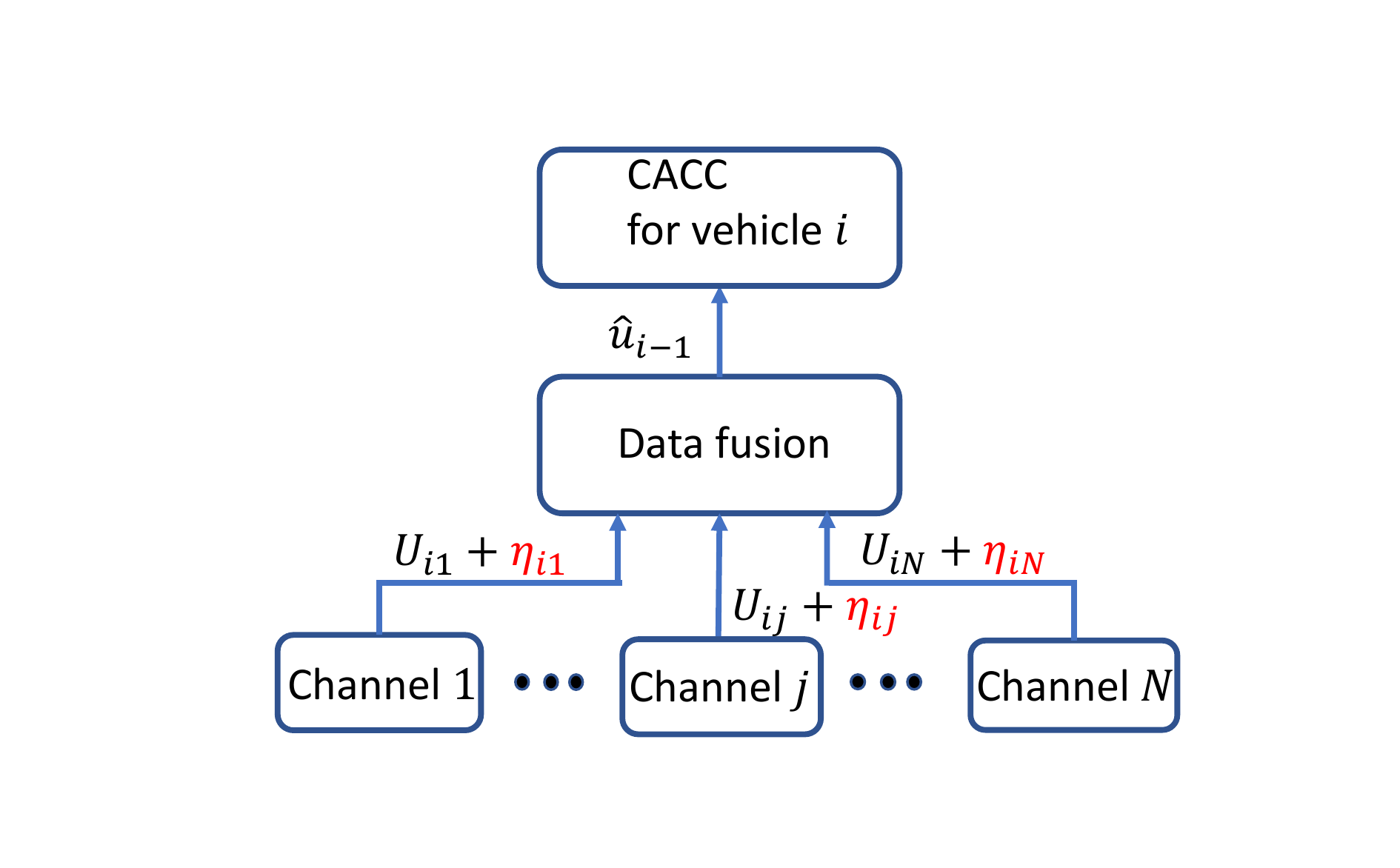}
		\caption{Schematic of the proposed data fusion algorithm at vehicle $i$.}
		\centering
		\label{fig:fff}
	\end{figure}
\begin{figure}[h]\centering
	\includegraphics[width=0.5\textwidth]{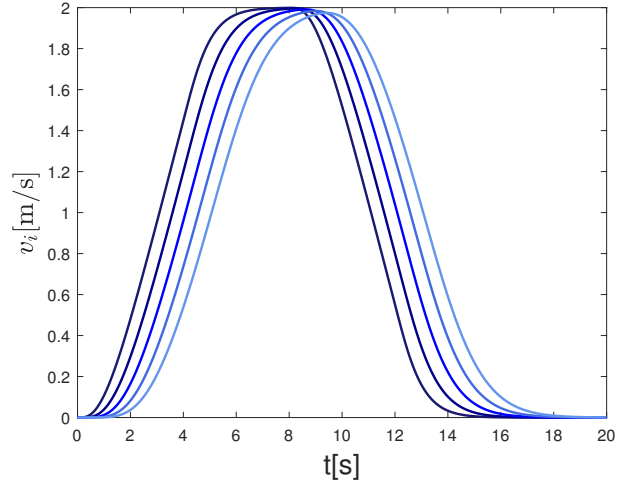}
	\caption{$H_{\infty}$ controller: measured velocity response at startup (black-light blue: vehicle 1-5).}
	\centering
	\label{fig:f4}
\end{figure}
\begin{figure}[h]\centering
	\includegraphics[width=0.5\textwidth]{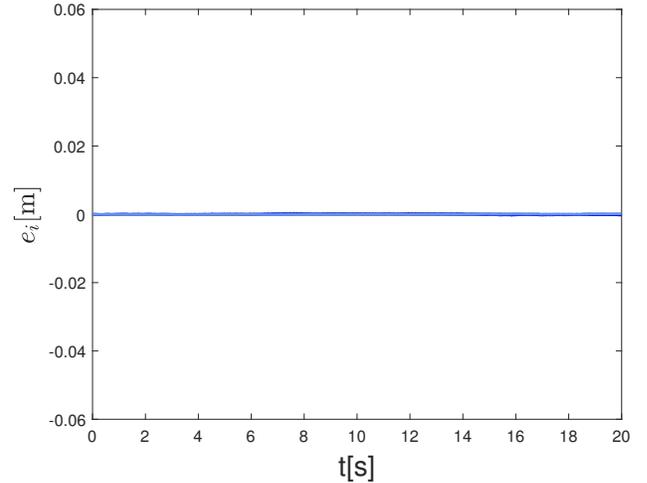}
	\caption{$H_{\infty}$ controller:  tracking error at startup (dark-light blue: vehicle 1-5).}
	\centering
	\label{fig:f5}
\end{figure}
\begin{figure}[h]\centering
	\includegraphics[width=0.5\textwidth]{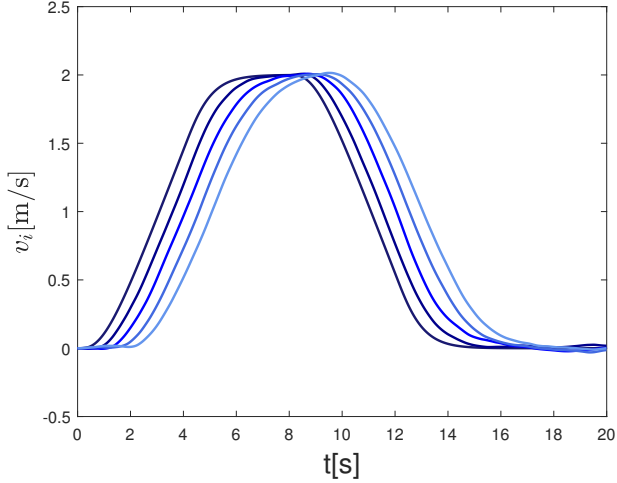}
	\caption{Controller given in \cite{Ploeg2014}:  measured velocity response at startup (dark-light blue: vehicle 1-5).}
	\centering
	\label{fig:f10}
\end{figure}
\begin{figure}[h]\centering
	\includegraphics[width=0.5\textwidth]{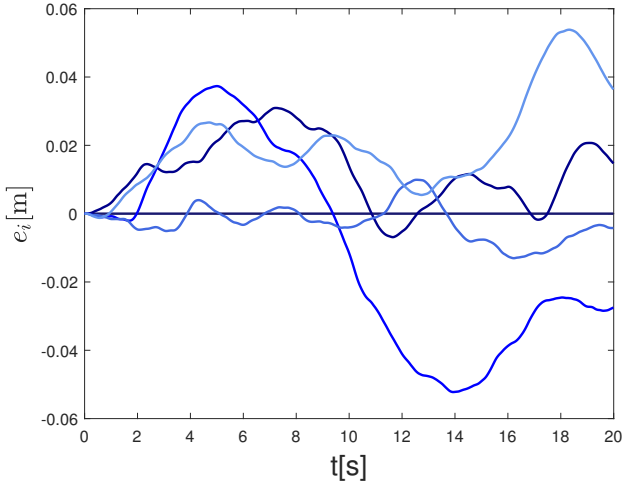}
	\caption{Controller given in \cite{Ploeg2014}, tracking error at startup (dark-light blue: vehicle 1-5).}
	\centering
	\label{fig:f11}
\end{figure}

	We have shown that the error of the proposed fusion algorithm is bounded independently of the attacks if the conditions of Theorem 2 are satisfied. In this section, we assume that the on-board sensors of each vehicle in the platoon are noisy as well and each vehicle uses the measurement provided by the fusion algorithm for control as shown in Figure \ref{fig:fff}. For each vehicle in the platoon, we design a $H_{\infty}$ controller to stabilize its closed-loop dynamics while minimizing the effect of the error introduced by the fusion algorithm and sensor noise on string stability and tracking performance.
	We consider system \eqref{s1} and assume the on-board sensors of vehicle $i$ that measure relative distance and relative velocity are subject to noise $\omega_{di}$ and $\omega_{vi}$, respectively. We assume a zero-order hold is used by vehicle $i$ between each sampling interval and let $\hat{u}_{i-1}(t)$ denote the converted continuous-time version of $\hat{u}_{i-1}(k)$, i.e., $\hat{u}_{i-1}(t)=\hat{u}_{i-1}(k)=u_{i-1}(k)+e_{\sigma_{i}(k)}(k)$ ($e_{\sigma_{i}(k)}(k)$ as defined in Theorem \ref{t2}) for $t\in[kT_{s},(k+1)T_{s})$, for all $i\in S_{m}$. Then, the closed-loop dynamics can be written as follows:
	\begin{equation}\label{e1}
		\begin{split}
			\begin{bmatrix}
				\dot{e}_{i}\\
				\dot{v}_{i}\\
				\dot{a}_{i}\\
				\dot{u}_{i}
			\end{bmatrix}=&\begin{bmatrix}
				0&-1&-h_{i}&0\\
				0&0&1&0\\
				0&0&-\frac{1}{\tau_{i}}&\frac{1}{\tau_{i}}\\
				\frac{k_{pi}}{h_{i}}&-\frac{k_{di}}{h_{i}}&-k_{d_{i}}&-\frac{1}{h_{i}}
			\end{bmatrix}\begin{bmatrix}
				{e}_{i}\\
				{v}_{i}\\
				{a}_{i}\\
				{u}_{i}
			\end{bmatrix}\\
			&+\begin{bmatrix}
				0&1&0\\
				0&0&0\\
				0&0&0\\
				\frac{k_{pi}}{h_{i}}&\frac{k_{d_{i}}}{h_{i}}&\frac{1}{h_{i}}
			\end{bmatrix}\begin{bmatrix}
				\omega_{di}\\
				v_{i-1}+\omega_{vi}\\
				u_{i-1}+e_{\sigma_{i}(k)}
			\end{bmatrix}.
		\end{split}
	\end{equation}
	Let $\omega_{i} := \begin{bmatrix}
		\omega_{di},v_{i-1}+\omega_{vi},u_{i-1}+e_{\sigma_{i}(k)}
	\end{bmatrix}^{\top}$. To implement the design method provided in \cite{he2006improved}, we rewrite the system in the following way:
	\begin{equation}\label{ss}
\left\{
		\begin{split}
			\dot{x}_{i}=&\tilde{A}_{i}x_{i}+\tilde{B}_{i1}w_{i}+\tilde{B}_{i2}\tilde{u}_{i},\\
			z_{i}=&\tilde{C}_{i1}x_{i},\\
			y_{i}=&\tilde{C}_{i2}x_{i}+\tilde{D}_{i21}w_{i},\\
			\tilde{u}_{i}=&K_{i}y_{i},
		\end{split}
\right.
	\end{equation}
	with
	\begin{equation}
\left\{
		\begin{split}
			\tilde{A}_{i}=&\begin{bmatrix}
				0&-1&h_{i}&0\\
				0&0&0&1\\
				0&0&-\frac{1}{\tau_{i}}&\frac{1}{\tau_{i}}\\
				0&0&0&-\frac{1}{h_{i}}
			\end{bmatrix},\tilde{C}_{i2}=\begin{bmatrix}
				1&0&0&0\\
				0&-1&-h_{i}&0
			\end{bmatrix},\\
			\tilde{B}_{i1}=&\begin{bmatrix}
				0&1&0\\
				0&0&0\\
				0&0&0\\
				0&0&\frac{1}{h_{i}}
			\end{bmatrix},\tilde{B}_{i2}=\begin{bmatrix}
				0\\0\\0\\\frac{1}{h_{i}}
			\end{bmatrix},			\tilde{D}_{i21}=\begin{bmatrix}
			1&0&0\\
			0&1&0
		\end{bmatrix},\\
			K_{i}=&\begin{bmatrix}
				k_{pi}&k_{di}
			\end{bmatrix},
		\end{split}
\right.
	\end{equation}
	where $z_{i}$ is the performance output we seek to optimize.  
	To minimize the effect of $\omega_{i}$ on string stability and tracking performance, we let $\tilde{C}_{i1}$ in \eqref{ss} be given as
	\begin{equation}\label{performance_matrix}
		\tilde{C}_{i1}=\begin{bmatrix}
			1&0&0&0\\
			0&1&0&0
		\end{bmatrix}.
	\end{equation}
This choice of $\tilde{C}_{i1}$ characterizes the performance output, $z_{i}=\tilde{C}_{i1}x$, and let us weight simultaneously the effect of $\omega_{i}$ on the tracking error, $e_i$, and the velocity of the vehicle, $v_i$, for string stability. We seek to synthesize the controller gains, $(k_{pi},k_{di})$, to minimize the $H_{\infty}$ norm from the vector of disturbances, $\omega_{i}$, to the performance output, $z_{i}=\tilde{C}_{i1}x$. To obtain such optimal gains, we use the reformulation in \eqref{e1}-\eqref{performance_matrix} and Algorithms 3 and 4 in \cite{he2006improved} constrained to satisfy $k_{pi}$, $k_{di}>0$, and $k_{di}>k_{pi}\tau_{i}$ to guarantee the vehicle following control objective \cite{Ploeg2014}. This emulation-based controller is able to stabilize \eqref{s1} when $T_{s}$ is sufficiently small \cite{tabbara2008networked}.\\[3mm]
	\textbf{Example 3:} We consider a homogeneous vehicle platoon consisting of $5$ vehicles. We let $h_{i}=0.5$, $\tau_{i}=0.1$, for all $i\in \left\lbrace 1,2,3,4,5\right\rbrace $, $T_{s}=0.01$ seconds, and $N_{i}=3$ communication channels between vehicles, $i\in\left\lbrace 2,3,4,5\right\rbrace$. Channel and sensor perturbations are taken as $\nu_{i1}\sim\mathcal{U}(-0.1,0.1)$, $\nu_{i2}\sim\mathcal{U}(-0.2,0.2)$, $\nu_{i3}\sim\mathcal{U}(-0.3,0.3)$, and $\omega_{di},\omega_{vi}\sim\mathcal{U}(-0.1,0.1)$. Let $\epsilon_{0}$ be given as in Table 1. At each time step $k$, let one of the $3$ communication channels between every two neighbor vehicles be randomly selected to be attacked, i.e., $W_{i}(k)=\left\lbrace 1\right\rbrace$, $\left\lbrace 2\right\rbrace $ or $\left\lbrace 3\right\rbrace $ and let $\eta_{W_{i}(k)}\sim\mathcal{N}(0,5^{2})$ for $i\in\left\lbrace 2,3,4,5\right\rbrace $. For all $k\geq 0$, the $i$-th vehicle $i\in\left\lbrace 2,3,4,5\right\rbrace $ uses \eqref{es1}-\eqref{es3} for fusing measurements of $u_{i-1}(k)$, and the CACC switching controller \eqref{1}-\eqref{3} to close the loop.\\
We use algorithms 3 and 4 in \cite{he2006improved} to design a robust controller for each vehicle in the platoon with smallest achievable $H_{\infty}$ norm of $\gamma=1.0198$, and corresponding optimal $k_{pi}=5.002$ and $k_{di}=305.1862$, for $i\in S_{m}$. The performance of the robust controller is shown in Figures \ref{fig:f4}-\ref{fig:f5}. For comparison, the performance of the controller in \cite{Ploeg2014} with $k_{pi}=0.2$, $k_{di}=0.7$ is shown in Figures \ref{fig:f10}-\ref{fig:f11}, where the $H_{\infty}$ gain from $\omega_{i}$ to $z_{i}$ is $5.1000$.

%
%
%

\subsection{Stability Analysis}
Here, we give a brief statement on the stability of the interconnected vehicles when the secure fused measurement given by \eqref{es1}-\eqref{es3} is used to drive the controller. Let $x_{i} = [e_{i},v_{i},a_{i},u_{i}]^{\top}$ and $\tilde{\epsilon}_{i} = [
		\omega_{di},v_{i-1}+\omega_{vi},u_{i-1}+e_{\sigma_{i}(k)}]^{\top}$. System \eqref{e1} can be written as:
	\begin{equation}\label{ee2}
		\dot{x}_{i}=A_{ci}x_{i}+\tilde{B}_{i}\tilde{\epsilon}_{i},
	\end{equation}
	with
	\begin{equation}
		A_{ci}=\begin{bmatrix}
			0&-1&-h_{i}&0\\
			0&0&1&0\\
			0&0&-\frac{1}{\tau_{i}}&\frac{1}{\tau_{i}}\\
			\frac{k_{pi}}{h_{i}}&-\frac{k_{di}}{h_{i}}&-k_{di}&-\frac{1}{h_{i}}
		\end{bmatrix},\tilde{B}_{i}=\begin{bmatrix}
			0&1&0\\
			0&0&0\\
			0&0&0\\
			\frac{k_{pi}}{h_{i}}&\frac{k_{di}}{h_{i}}&\frac{1}{h_{i}}
		\end{bmatrix}.
	\end{equation}
The closed-loop dynamics \eqref{ee2} is Input-to-State-Stable (ISS) with input $\tilde{\epsilon}_{i}$ because $A_{ci}$ is a Hurwitz matrix \cite{sontag2008input}. As we have proved in Theorem 2 that $e_{\sigma_{i}}(k)$ is bounded for all $k\geq 0$, if $||x_{i-1}||_{\infty}$ is bounded, then $||\tilde{\epsilon}_{i}||_{\infty}$ is bounded, which then implies the boundedness of $||x_{i}||_{\infty}$ since \eqref{ee2} is ISS. Therefore, the boundedness of $||x_{i-1}||_{\infty}$ implies the boundedness of $||x_{i}||_{\infty}$ for $i\in S_{m}$. From the fact that the state of the virtual referece vehicle is bounded, i.e., $||x_{0}||_{\infty}$ is bounded, we conclude the boundedness of $||x_{i}||_{\infty}$ for all $i\in S_{m}$ \cite{sontag2008input}.


	\section{Conclusion}\label{conclusion}
	We have addressed the problem of data fusion, attack detection and isolation, and robust control for connected vehicles whose communication channels are under (potentially unbounded) cyber attacks. We suggest creating redundancy of inter-vehicle communication channels for connected vehicles to enhance their resilience to cyber attacks so that degradation from CACC to ACC can be avoided. Exploiting network redundancy, we have proposed a data fusion framework that reconstructs the transmitted data via the networks with bounded errors independent of attacks on the communication networks. These fused information is then used to detect and isolate attacks and stabilize the closed-loop dynamics of each vehicle. An $H_{\infty}$ controller is designed for each vehicle in a platoon to stabilize its closed-loop dynamics while minimizing the effects of sensor noise and fusion errors on string stability and tracking performance.
\appendices
\section{Proof of Theorem 1}\label{the1}
\textbf{$(i)\to (ii)$}
We proceed by contraposition. Assume $q_{i}\geq\frac{N_{i}}{2}$, then for all $W_{i}(k)\subset\left\lbrace 1, \ldots, N_{i}\right\rbrace $, with $\card(W_{i}(k))=q_{i}$ and $k\geq 0$, there exists another set $\bar{W}_{i}(k)\subset\left\lbrace 1, \ldots, N_{i}\right\rbrace $ with $\card(\bar{W}_{i}(k))=q_{i}$ such that $W_{i}(k)\cup \bar{W}_{i}(k)=\left\lbrace 1, \ldots, N_{i}\right\rbrace $ since $2q_{i}\geq N_{i}$. Let $I_{i}(k)=W_{i}(k)\cap \bar{W}_{i}(k)$. For all $\bar{u}_{i-1}(k) \neq u_{i-1}(k)$, with corresponding $\bar{U}_i(k) = \mathbf{1}\bar{u}_{i-1}(k) + \nu_i(k) + \bar{\eta}_i(k)$ and $U_i(k) = \mathbf{1}u_{i-1}(k) + \nu_i(k) + \eta_i(k)$, we seek $\bar{\eta}_i(k)$ and ${\eta}_i(k)$ such that $\bar{U}_i(k) = U_i(k)$. If such attack vectors exist, we can conclude that $q_{i} \geq \frac{N_{i}}{2}$ implies $u_{i-1}(k)$ is not reconstructible from $U_i(k)$ (according to Definition 1), and thus, by contraposition, $q_{i} < \frac{N_{i}}{2}$ implies $u_{i-1}(k)$ is reconstructible \cite{loehr2019introduction}. For $\bar{u}_{i-1}(k) \neq u_{i-1}(k)$, let
\begin{equation*}
	j\in \left\lbrace \bar{W}_{i}(k)\setminus I_{i}(k)\right\rbrace:\left\{
	\begin{split}
		&\bar{\eta}_{ij}(k)=u_{i-1}(k)-\bar{u}_{i-1}(k),\\ &\eta_{ij}(k)=0, 
	\end{split}
	\right.
\end{equation*}
\begin{equation*}
	j\in \left\lbrace W_{i}(k)\setminus I_{i}(k)\right\rbrace:\left\{
	\begin{split}
	&\bar{\eta}_{ij}(k)=0,\\
	& \eta_{ij}(k)=\bar{u}_{i-1}(k)-u_{i-1}(k),
	\end{split}
	\right.
\end{equation*}
\begin{equation*}
	j\in I_{i}(k).:\left\{
	\begin{split}
&\bar{\eta}_{ij}(k)=u_{i-1}(k),\\
&\eta_{ij}(k)=\bar{u}_{i-1}(k).
	\end{split}
	\right.
\end{equation*}
It is easy to verify that these particular $\bar{\eta}_i(k)$ and ${\eta}_i(k)$ satisfy  $u_{i-1}(k)\neq \bar{u}_{i-1}(k)$ and $U_{i}(k)=\bar{U}_{i}(k)$, for all $k\geq 0$, and the result follows.\\

\textbf{$(ii)\to (i)$}
Let $q_{i}<\frac{N_{i}}{2}$. For all $\eta_{i}(k)$ and $\bar{\eta}_{i}(k)$ with $\card(\supp(\eta_{i}(k)))\leq q_{i}$ and $\card(\supp(\bar{\eta}_{i}(k)))\leq q_{i}$, the number of nonzero elements of the difference, $\bar{\eta}_{i}(k)-\eta_{i}(k)$, are less than or equal to $2q_{i}$, which is strictly less than $N_{i}$. Therefore, there does not exist $u_{i-1}(k) \neq \bar{u}_{i-1}(k)$ such that
\begin{equation}\label{p1}
	\bar{\eta}_{i}(k)-\eta_{i}(k)=\left[ \begin{matrix}
		u_{i-1}(k)-\bar{u}_{i-1}(k)\\
		u_{i-1}(k)-\bar{u}_{i-1}(k)\\
		\vdots\\
		u_{i-1}(k)-\bar{u}_{i-1}(k)
	\end{matrix}\right],
\end{equation}
because the number of nonzero elements on the right side of \eqref{p1} has to be strictly less than $N_{i}$, which indicates that there does not exist $u_{i-1}(k)\neq\bar{u}_{i-1}(k)$ such that $U_i(k) = \bar{U}_i(k)$, i.e.,
\begin{equation}\label{eq1}
	\left[ \begin{matrix}
		u_{i-1}(k)+\nu_{i1}(k)\\
		u_{i-1}(k)+\nu_{i2}(k)\\
		\vdots\\
		u_{i-1}(k)+\nu_{iN}(k)
	\end{matrix}\right] +\eta_{i}(k)=\left[ \begin{matrix}
		\bar{u}_{i-1}(k)+\nu_{i1}(k)\\
		\bar{u}_{i-1}(k)+\nu_{i2}(k)\\
		\vdots\\
		\bar{u}_{i-1}(k)+\nu_{iN}(k)
	\end{matrix}\right] +\bar{\eta}_{i}(k).
\end{equation}
Hence, $u_{i-1}(k)$ is reconstructible from $U_{i}(k)$ under $q_{i}$ attacks (according to Definition 1).
\section{Proof of Lemma 1}\label{l1}
By construction and the triangle inequality
\begin{equation}
	\begin{split}
		&\left|\sum_{j=1}^{j=n}\nu_{ij}\right| = \sqrt{(\nu_{i1}+\nu_{i2}+\cdots+\nu_{in})^{2}}\\
		=&\sqrt{\nu_{i1}^{2}+\nu_{i2}^{2}+\cdots+\nu_{in}^{2}+2(\nu_{i1}\nu_{i2}+\cdots+\nu_{i(n-1)}\nu_{in})}\\
		\leq&\sqrt{\nu_{i1}^{2}+\nu_{i2}^{2}+\cdots+\nu_{in}^{2}+(\nu_{i1}^{2}+\nu_{i2}^{2}+\cdots+\nu_{i(n-1)}^{2}+\nu_{in}^{2})}\\
		=&\sqrt{\left( 1+\left( \begin{matrix}
				1\\
				n-1
			\end{matrix}\right) \right) \left( \nu_{i1}^{2}+\nu_{i2}^{2}+\cdots+\nu_{in}^{2}\right\rbrace }\\
		\leq&\sqrt{n^{2}||\nu_{i}||_{\infty}^{2}}\\
		=&n||\nu_{i}||_{\infty}.
	\end{split}
\end{equation}
It follows that
\begin{equation}
	\begin{split}
		|\hat{u}_{iJ}(k)-u_{i-1}(k)|=&\left| \frac{\sum_{j\in J}U_{ij}(k)}{\card(J)}-u_{i-1}(k)\right|\\
		=&\frac{1}{\card(J)}\left|\sum_{j\in J}\nu_{ij}(k)\right|\\
		{\leq}&\frac{1}{\card(J)}\card(J)||\nu_{i}||_{\infty}\\
		=&||\nu_{i}||_{\infty}.
	\end{split}
\end{equation}
\section{Proof of Theorem 2}\label{the2}
	Under Assumption \ref{a1}, there exists at least one subset $\bar{I}(k)\subset\left\lbrace 1, \ldots, N_{i}\right\rbrace $ with $\card(\bar{I}(k))=N_{i}-q_{i}$ such that $\eta_{i}^{\bar{I}(k)}(k)= \mathbf{0}$ for $k\geq 0$. Then
\begin{equation}
	|\hat{u}_{\bar{I}(k)}(k)-u_{i-1}(k)|\leq ||\nu_{i}||_{\infty}.
\end{equation}
Moreover, for all $j\in\bar{I}(k)$, $\eta_{ij}(k)=0$ and we have
\begin{equation}
	|U_{ij}(k)-u_{i-1}(k)|=|\nu_{ij}(k)|.
\end{equation}
Then, we have
\begin{equation}
	\begin{split}
		\pi_{i\bar{I}(k)}(k)=&\underset{j\in \bar{I}(k)}{\max}\left| \hat{u}_{\bar{I}(k)}(k)-U_{ij}(k)\right| \\
		=&\underset{j\in \bar{I}(k)}{\max}\left|\hat{u}_{\bar{I}(k)}(k)-u_{i-1}(k)+u_{i-1}(k)-U_{ij}(k)\right| \\
		=&\left| \hat{u}_{\bar{I}(k)}(k)-u_{i-1}(k)\right| +\underset{j\in \bar{I}(k)}{\max}\left| u_{i-1}(k)-U_{ij}(k)\right| \\
		\leq&||\nu_{i}||_{\infty}+\underset{j\in \bar{I}(k)}{\max}|\nu_{ij}(k)|.
	\end{split}
\end{equation}
From Corollary \ref{c2}, among every $N_{i}-q_{i}$ communication channels, at least one of the channels is attack free since $N_{i}-2q_{i}\geq 1$. Therefore, there exists at least one $\bar{j}(k)\in\sigma_{i}(k)$ such that $\eta_{i\bar{j}(k)}(k)=0$ for $k\geq 0$ and
\begin{equation}
	\begin{split}
		|U_{i\bar{j}}(k)-u_{i-1}(k)|=|\nu_{i\bar{j}(k)}(k)|.
	\end{split}
\end{equation}
From \eqref{es2}, we have $\pi_{i\sigma_{i}(k)}(k)\leq\pi_{i\bar{I}(k)}(k)$. From \eqref{es1}, we have
\begin{equation}
	\begin{split}
		\pi_{i\sigma_{i}(k)}(k)=&\underset{j\in\sigma_{i}(k)}{\max}\left| \hat{u}_{\sigma_{i}(k)}(k)-U_{ij}(k)\right| \\
		\geq&\left| \hat{u}_{\sigma_{i}(k)}(k)-U_{i\bar{j}(k)}(k)\right|.
	\end{split}
\end{equation}
Using the lower bound on $\pi_{i\sigma_{i}(k)}(k)$ and the triangle inequality, we have that
\begin{equation}\label{ee}
	\begin{split}
		|e_{\sigma_{i}(k)}(k)|=&|\hat{u}_{\sigma_{i}(k)}-u_{i-1}(k)|\\
		=&\left| \hat{u}_{\sigma_{i}(k)}(k)-U_{i\bar{j}(k)}(k)+U_{i\bar{j}(k)}(k)- u_{i-1}(k)\right| \\
		\leq&\pi_{\sigma_{i}(k)}(k)+|\nu_{i\bar{j}}(k)|\\
		\leq&\pi_{i\bar{I}(k)}(k)+|\nu_{i\bar{j}}(k)|\\
		\leq&||\nu_{i}||_{\infty}+\underset{j\in \bar{I}(k)}{\max}|\nu_{ij}(k)|+|\nu_{i\bar{j}}(k)|\\
		\leq&3 ||\nu_{i}||_{\infty}.
	\end{split}
\end{equation}
Inequality \eqref{ee} is of the form \eqref{sa} and the result follows.
	\bibliographystyle{ieeetr}
	\bibliography{Observer1}
\end{document}